\documentclass[10pt]{iopart}
\usepackage{iopams}
\usepackage{graphicx}

\usepackage{braket}
\usepackage{upgreek}
\usepackage{subscript}
\usepackage{lmodern}
\usepackage{color}
\expandafter\let\csname equation*\endcsname\relax

\expandafter\let\csname endequation*\endcsname\relax

\usepackage{amsmath}
\usepackage{amsfonts}
\usepackage{amssymb}

\newcommand{\rubidium}{\textsuperscript{87}Rb}
\newcommand{\us}{\,$\upmu$s}
\newcommand{\um}{\,$\upmu$m}

\newcommand{\insitu}{\,$\textit{in-situ}$}
\newcommand{\rbohr}{\textit{a}\textsubscript{0}}
\newcommand{\density}[2]{#1$\times$10\textsuperscript{#2}\unit{}{cm\textsuperscript{-3}}}

\newcommand{\vecr}{\vec{r}}
\newcommand{\RydGS}{\textit{nS}\,+\,5\textit{S}}
\newcommand{\RydGD}{\textit{nD}\,+\,5\textit{S}}
\newcommand{\RydGP}{\textit{nP}\,+\,5\textit{S}}

\newcommand{\unit}[2]{\mbox{#1\,#2}}

\newcommand{\MHz}{\,MHz}
\newcommand{\kHz}{\,kHz}
\newcommand{\Hz}{\,Hz}
\newcommand{\nK}{\,nK}
\newcommand{\cm}{\,cm}
\newcommand{\mm}{\,mm}
\newcommand{\nm}{\,nm}
\newcommand{\Volt}{\,V}
\newcommand{\becatoms}{1.5$\times$10\textsuperscript{6}\unit{}}

\begin{document}

\title{Controlling Rydberg atom excitations in dense background gases }
\author{Tara Cubel Liebisch, Michael~Schlagm\"{u}ller, Felix Engel, Huan Nguyen, Jonathan Balewski, Graham Lochead, Fabian B\"{o}ttcher, Karl M. Westphal, Kathrin S. Kleinbach, Thomas Schmid, Anita Gaj, Robert L\"{o}w, Sebastian Hofferberth, and Tilman Pfau} 
\address{5. Physikalisches Institut and Center for Integrated Quantum Science and Technology IQST, Universit\"at Stuttgart, Pfaffenwaldring 57, 70569 Stuttgart, Germany}
\author{Jes\'{u}s P\'{e}rez-R\'{i}os and Chris H. Greene}
\address{Department of Physics and Astronomy, Purdue University, West Lafayette, IN, 47907 USA}
\ead{pfau@physik.uni-stuttgart.de}
\vspace{10pt}
\begin{indented}
\item[]{\today}
\end{indented}

\begin{abstract}
We discuss the density shift and broadening of Rydberg spectra measured in cold, dense atom clouds in the context of Rydberg atom spectroscopy done at room temperature, dating back to the experiments of Amaldi and Segr\`e in 1934.  We discuss the theory first developed in 1934 by Fermi to model the mean-field density shift and subsequent developments of the theoretical understanding since then.  In particular, we present a model whereby the density shift is calculated using a microscopic model in which the configurations of the perturber atoms within the Rydberg orbit are considered.  We present spectroscopic measurements of a Rydberg atom, taken in a Bose-Einstein condensate (BEC) and thermal clouds with densities varying from \density{5}{14} to \density{9}{12}.  The density shift measured via the spectrum's center of gravity is compared with the mean-field energy shift expected for the effective atom cloud density determined via a time of flight image.  Lastly, we present calculations and data demonstrating the ability of localizing the Rydberg excitation via the density shift within a particular density shell for high principal quantum numbers.

\end{abstract}


%

%

\section{Introduction}\label{introduction}
\subsection{Rydberg atom excitations in cold, dense media}\label{densemedia}
\begin{figure}
	\centering
		\includegraphics[width=0.9\textwidth]{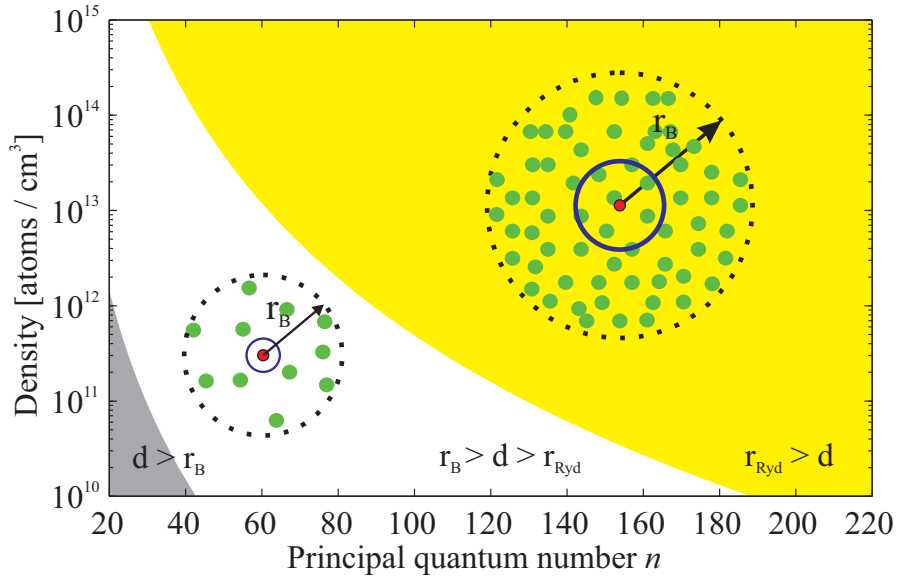}
	\caption{(color online) Plot of density versus principal quantum number showing three regimes.  At the lowest \textit{n}'s and lowest densities (gray), the interparticle spacing, d, is greater than the Rydberg blockade radius, r$_{\text{B}}$, assuming an excitation bandwidth of 1\MHz.  In the intermediate regime (white), the Rydberg blockade radius is larger than the interparticle spacing, but there are no background atoms overlapped with the Rydberg orbit with a radius of r$_{\text{Ryd}}$.  In the third regime (yellow), at the highest \textit{n}'s and highest densities, many background atoms are contained within the Rydberg blockade sphere and the Rydberg orbit. 
	}
	\label{Fig1:RydbergvsDensity}
\end{figure}

If a Rydberg atom is excited in a background gas where there is a non-negligible probability that one or more perturber atoms are located within the Rydberg electron orbit, then it is considered to be excited in a high density background gas.  This definition of high density Rydberg excitation implies that an inter-dependence of the chosen principal quantum number, \textit{n}, and background gas density, dictates whether or not this regime is realized.  The number of background perturber atoms within the Rydberg orbit can be tuned between zero atoms and tens of thousands of atoms in typical experiments~\cite{Gaj2014}, as shown in Figure~\ref{Fig1:RydbergvsDensity}.  An understanding of Rydberg spectroscopy and the limits of the collisional lifetimes in dense gases is particularly important as typical Rydberg atom experiments are pushed to higher density regimes, in order to maximize the Rydberg blockade effect, and thereby prevent excitation for a large amount of other atoms~\cite{Saffman2010}.  Examples include studies of quantum information~\cite{Saffman2010}, single photon transistors~\cite{Gorniaczyk2014, Tiarks2014}, and all optical quantum processing~\cite{Paredes2014}.  Schemes to realize Rydberg-based phase gates, utilizing optical cavities, have been proposed in order to circumvent undesired limitations of the Rydberg collisional lifetime in high density gases~\cite{Das2015}. 

A Rydberg atom impurity excited in a dense background gas is a testbed of fundamental quantum mechanical electron-perturber scattering for a range of low electron energies (0\,-\,30\,meV) not achievable in electron beams~\cite{Hordon1993}.  Spectroscopy of Rydberg atoms in dense background gases has a long history that dates back to the work of Amaldi and Segr\`e carried out in 1934~\cite{Amaldi1934}.  In their work they studied the influence of high density background gases such as hydrogen, nitrogen, helium and argon on the Rydberg absorption spectra of sodium and potassium, with background gas densities as high as \density{3}{19}, whereby more than 10,000 perturber atoms were located within the classical Rydberg orbit.  They observed, somewhat surprisingly, that the Rydberg atom could be excited in such a dense background gas.  The line shift and broadening of the absorption spectra depended on the density and species of the background atoms.  A theoretical framework was developed by Fermi in 1934 to explain the shifted spectral lines~\cite{Fermi1934}.  It was determined that the Rydberg electron scattering with background perturber atoms dominated the observed spectral line shifts.  The derivation by Fermi of the mean-field density shift will be explained in detail in section~\ref{Pseudopotential}.  

A renewed interest in high-density Rydberg spectroscopy took place starting in the 1960's in which theoretical understanding of fundamental electron-perturber scattering beyond the \textit{s}-wave limit of the Fermi pseudopotential was developed~\cite{Alekseev1966, Presnyakov1970, Omont1977}.  The experimental work of this time period explored a new regime, namely Rydberg spectroscopy with alkali atom background atoms.  These studies were performed in thermal vapor cells with densities reaching as high as \density{1}{15}~\cite{Stoicheff1980, Thompson1987}.
This work inspired new theoretical studies in order to explain the \textit{n}-dependent line shift and line broadening observed in electron-perturber alkali atom collisions.   In particular, this led to the prediction of electron-neutral scattering shape resonances for \textit{p}-wave scattering between an electron and an alkali atom~\cite{Fabrikant1986, Borodin1991}, which was later confirmed via \textit{ab initio} calculations~\cite{Bahrim2000}.  A discussion of these theoretical developments can be found in~\cite{Khuskivadze2002}.  

A third major period for studying Rydberg atoms in dense media is currently occurring due to the advancement of laser trapping and cooling.  Rydberg spectroscopy performed in background gases with densities equivalent to the previous room temperature experiments can be achieved, but now at thermal energies eight orders of magnitude lower.  This new experimental regime of cold, dense background perturber atoms has led to new discoveries of the role of fundamental electron-neutral interactions in ultracold chemistry.  This is because the cold temperatures enable high-resolution Rydberg atom spectroscopy.  In particular, the interaction between the Rydberg electron and the neutral atom perturber leads to a potential well ranging in depth from kHz to GHz, depending on the wavefunction of the Rydberg electron and the scattering phase shifts~\cite{Greene2000, Chibisov2002, Hamilton2002}.  Exploiting these two properties, ultra-long range Rydberg molecules have been observed, as first predicted in 2000 by Greene et al.~\cite{Greene2000}.  Stable Rydberg molecules with large internuclear separations have been observed for \textit{S}, \textit{P}, and \textit{D} states for Rb, Cs, and Sr~\cite{Bendkowsky2009, Butscher2011, Gaj2014, Gaj2015, Tallant2012, DeSalvo2015, Anderson2014, Krupp2014, Bellos2013}, with large dipole moments~\cite{Li2011, Booth2015}, and with ground states bound by a mixture of singlet and triplet electron-neutral scattering~\cite{Sassmannhausen2015, Boettcher2015}. 
	
In this review we discuss the fundamental Rydberg electron-neutral interactions critical to understanding Rydberg spectroscopy in cold, dense media.  In particular, we build on the point-particle perturber model, presented in~\cite{Schlagmueller2015}, to demonstrate that the configurations of background perturbing atoms, with respect to the potential energy curves of the electron-perturber system, must be considered in order to not only understand the spectroscopic line broadening as discussed in~\cite{Schlagmueller2015}, but also the line shift.  This model demonstrates deviations from the Fermi mean-field model in estimating the spectroscopic line shift.  The discussion of the density shift outlined in this review, which goes beyond Fermi's density shift model, should serve as an introduction to the idea that large deviations from the mean-field density shift are expected for various combinations of the Rydberg quantum state and the electron-perturber spin orientation.

We discuss in detail in this review to what extent Rydberg spectroscopy in cold, dense samples can be used to probe the local density, temperature, and perhaps even the density correlations of a cold atom sample.  The ability to measure the local density distribution of a system is a key aspect to studying interacting many-body systems in ultracold quantum gases.   One prominent example is that the mean-field interaction alters the ground state of a harmonically trapped Bose gas from a Gaussian distribution to a Thomas-Fermi distribution~\cite{Pitaevskii2003}.  While time-of-flight absorption imaging~\cite{Ketterle1999} has been a long-established standard to extract information about the dynamics and the properties of a quantum gas, \insitu~imaging can reveal important information about the system, lost in time of flight~\cite{Ku2012, Duarte2015, Gericke2008, Heinze2013, Reinhard2013, Nascimbne2010, Sherson2010, Sadler2006, Gemelke2009, Murthy2014}.  In these experiments phase-contrast, fluorescence, and dark-ground imaging in the optical domain, as well as electron beam imaging have been used to measure \insitu~properties of the quantum gas.  Recent experiments have even achieved \insitu~measurements of the local gas density~\cite{Ku2012, Duarte2015, Reinhard2013, Nascimbne2010, Manthey2015}, and interesting new non-destructive density probes have been proposed~\cite{Hangleiter2015}.  In this work we discuss how Rydberg atom spectroscopy can be used as a probe of the local density and temperature.  The size of a Rydberg atom scales as $n^{*2}$, where $n^*$ is the effective principal quantum number taking the quantum defect $\delta_l$ into account.  Therefore, the local probe size can, in principle, be tuned from 100's of~\nm s~to~\um~sizes, by exciting Rydberg states from $n=50-200$.

\section{Rydberg electron-neutral scattering}\label{PECs}
\subsection{The Fermi pseudopotential}\label{Pseudopotential}
The experimentally observed spectroscopic line shift of Rydberg atoms excited in dense background gases, is dominated by low energy electron-perturber scattering events. Fermi's model explicitly considers the Rydberg electron as a quasi-free particle that is scattered with low momentum $\hbar k$ from the ground state atoms.  This approach was further refined by extending it to higher momenta~\cite{Alekseev1966} and alkali metal perturber atoms~\cite{Presnyakov1970}, recovering the original result, as an approximation, in the limit of \textit{s}-wave scattering. An excellent review of the extensions of the Fermi model can be found in~\cite{Omont1977}.  In this section the original derivation of Fermi's pseudopotential~\cite{Fermi1934,Reinsberg1934} is outlined.  This derivation is unconventional for modern textbook nomenclature~\cite{LL}, but it is very instructive for gaining a deeper understanding of the density shift and for gaining a perspective of the advances in the theoretical understanding of the density shift.

\begin{figure}
	\centering
		\includegraphics[width=1\columnwidth]{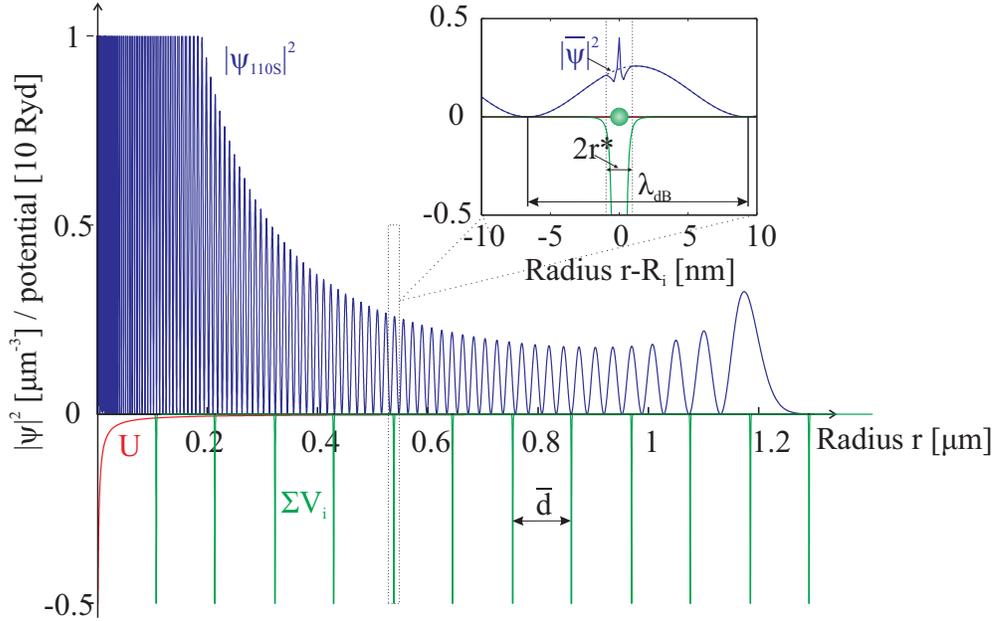}
	\caption{(color online) Semi-quantitative sketch of length scales: Radial profile of probability density $|\psi_{110S}|^2$ for an electron in the $110S$ Rydberg state. The Coulomb potential $U(r)$, created by the positively charged Rydberg core (red) and the polarization potential $\sum_{i}V_i$ by neutral $^{87}$Rb ground state atoms with polarizability $\alpha$\,=\,$4\pi\epsilon_0\cdot4.74\times10\textsuperscript{-29}\unit{}\,\text{m}^3$ (green)~\cite{Gregoire2015}, are indicated. Locally, the wavefunction of the Rydberg electron is deformed by the polarization potential. Averaging over one interaction range, given as the characteristic radius $r^*\approx0.96\,\text{nm}$, leads to the dashed probability density~$|\overline{\psi}|^2$ of the Rydberg electron, as indicated in the inset. The nearest neighbor spacing of $\overline{d}=110\,\text{nm}$ corresponds to an atomic density of \density{2}{14}. The indicated value of the de Broglie wavelength $\lambda_{\text{dB}}$ is just providing an order of magnitude, since it varies with position. 
	}
	\label{fig:Fermischeme}
\end{figure}

The starting point is the stationary Schr\"odinger equation for the Rydberg electron in the combined potential, created by the Rydberg core $U(\vec{r})=-\frac{e^2}{4\pi\epsilon_0}\frac{1}{|\vec{r}|}$ and the surrounding neutral ground state atoms $\sum_{i}V_i$:
\begin{equation} \label{eq:FermiSGl}
	\left[-\frac{\hbar^2}{2m_e}\Delta+U(\vec{r})+\sum_i V_i\right]\psi(\vec{r})=E\psi(\vec{r})
\end{equation}
The interaction potential~$V_i$ is assumed to be short-range and isotropic. This is the case, since the interaction is given by the polarization of an atom at position~$\vec{R}_i$ with polarizability $\alpha$~\footnote{SI units are used throughout section~\ref{PECs}.  The SI Units of polarizability are C$^2$m$^2$J$^{-1}$ as defined by the National Institute of Standards and Technology (NIST).  This is not to be confused with the cgs units for polarizability expressed as cm$^3$, and typically referred to as a polarizability volume.}:
\begin{equation} \label{eq:Vpol}
	V_i=-\frac{1}{(4\pi\epsilon_0)^2}\frac{\alpha e^2}{2|\vec{R}_i-\vec{r}|^4}
\end{equation}
The range of the polarization potential can be estimated by the characteristic radius $r^*$. This is the distance, at which the polarization potential equals the centrifugal potential energy, $\hbar^2/(2\mu r^2)$, with $\mu$ denoting the reduced mass of the electron and the perturber atom \cite{Zipkes2010}:
\begin{equation}
	r^{*}=\frac{\sqrt{\mu\alpha e^2}}{4\pi\epsilon_0\hbar}.
\end{equation}     
For ground state $^{87}$Rb, the characteristic radius is $r^{*}=0.96$\,\nm, which is much smaller than the nearest neighbor distance of $\overline{d}=110$\,\nm~for a density of \density{2}{14}.  Furthermore, the Rydberg electron is assumed to be slow; in particular, the de Broglie wavelength $\lambda_{\text{dB}}$ is required to be much larger than the interaction range $\sim r^{*}$. In Figure~\ref{fig:Fermischeme} the relevant length scales are sketched for the $110S$ Rydberg state. The assumptions are well justified, except for regions very close to the Rydberg core, where the de Broglie wavelength $\lambda_{\text{dB}}$ becomes small. 
 
\begin{figure}
	\centering
		\includegraphics[width=0.9\textwidth]{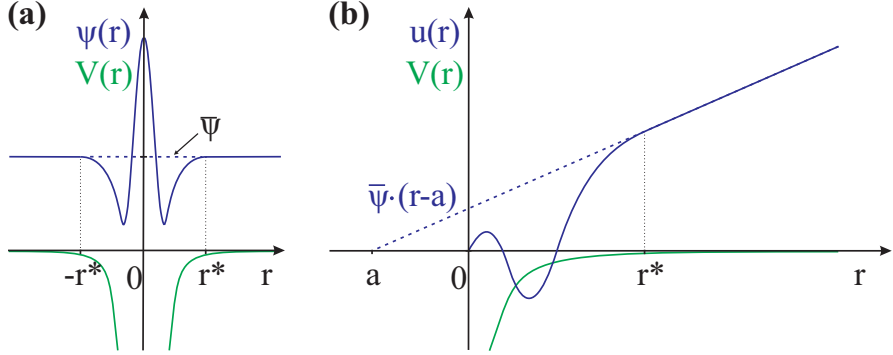}
	\caption{(color online) Sketched behavior of the Rydberg electron wavefunction $\psi$(r) (a) and r$\psi$(r) (b) close to a perturbing atom. The region of interest is much smaller than the de Broglie wavelength $\lambda_{\text{dB}}$, so that outside the interaction zone ($r>r^*$), the wavefunction becomes constant $\psi(r)=\overline{\psi}$.
	}
	\label{fig:FermischemeWF}
\end{figure}

The Schr\"odinger equation~(\ref{eq:FermiSGl}) is then averaged over a small volume. This volume is chosen smaller than the de Broglie wavelength $\lambda_{\text{dB}}$, but larger than the range $\sim r^*$ of the polarization potential. The averaged wavefunction~$\overline{\psi}(\vec{r})$ outside the interaction region then resembles the wavefunction $\psi(\vec{r})$. In this way the local impact of the perturbing neutral atoms is removed (see inset of Figure~\ref{fig:Fermischeme}). Under this condition, the mean of the derivative of the wavefunction can be replaced by the derivative of the averaged wavefunction ($\overline{\Delta\psi}=\Delta\overline{\psi}$) and one obtains:
\begin{equation} \label{eq:FermiSGlav}
	-\frac{\hbar^2}{2m_e}\Delta\overline{\psi}(\vec{r})-\left(E-U(\vec{r})\right)\overline{\psi}(\vec{r})+\overline{\sum_i V_i\psi(\vec{r})}=0.
\end{equation}
Now, the region with distance $r=|\vec{r}-\vec{R}_i|$, $0<r<\lambda_{\text{dB}}$, closely around a single perturbing neutral atom, is considered. This volume extends over regions with vanishing and non-vanishing interaction $V(r)$ with the perturber. Outside of the interaction zone ($r>r^*$), the wavefunction~$\psi(\vec{r})$ is equal to the averaged wavefunction $\overline{\psi}(\vec{r})$ and to first order constant, since~$r$ is smaller than the de Broglie wavelength $\lambda_{\text{dB}}$. Inside the interaction zone ($r\leq r^*$), the energy $E-U(\vec{r})$ is much smaller than the interaction with the perturbing atom, $V(r)$, and can be neglected. Therefore, the wavefunction $\psi(\vec{r})$ in the considered region becomes radially symmetric around the perturbing ground state atom and reads (see Figure~\ref{fig:FermischemeWF}a):
\begin{equation} \label{eq:FermiWF1}
	\psi(\vec{r})=\left\{\begin{array}{ll} \overline{\psi}=const. & \text{for} \ r>r^* \\
																				\psi(r) & \text{for} \ r\leq r^*. \end{array}\right.
\end{equation}
The Schr\"odinger equation~(\ref{eq:FermiSGl}) inside the interaction zone $r\leq r^*$ then simplifies to:
\begin{equation} \label{eq:FermiSGlWW}
\frac{1}{r^2}\frac{\partial}{\partial r}\left(r^2\frac{\partial\psi}{\partial r}\right) =\frac{2m_e}{\hbar^2}V(r)\psi(r).
\end{equation}
Considering the Laplace operator in spherical coordinates, a more suitable form can be obtained using the substitution 
\begin{equation} \label{eq:FermiSubs}
	\psi(r)=u(r)/r ,
\end{equation}
which then leads to:
\begin{equation} \label{eq:FermiSGlWWu}
	\frac{\partial^2u}{\partial r^2} =\frac{2m_e}{\hbar^2}V(r)u(r).
\end{equation}
For large distances $r$, the interaction potential~$V(r)$ vanishes and $u(r)$ thus becomes linear, as $r$ is approaching $r^*$. The slope of $u(r)$ is fixed by equation~(\ref{eq:FermiWF1}) to $\overline{\psi}$, so that $u(r)$ reads:
\begin{equation} \label{eq:FermiWF2}
	u(r)=\overline{\psi}\cdot(r-a) \ \ \text{for} \ r>r^*.
\end{equation}
Here, $a$ is the intercept of the asymptotic wavefunction, $\overline{\psi}$,~(\ref{eq:FermiWF2}) with the abscissa (see Figure~\ref{fig:FermischemeWF}b). This quantity is the \textit{s}-wave scattering length~\footnote{The sign here is chosen different to Fermi \cite{Fermi1934} in order to match modern textbook convention~\cite{LL}.}
.
Now, the averaged interaction energy stemming from one single perturber $i$ is calculated using the substitution~(\ref{eq:FermiSubs}) and equation~(\ref{eq:FermiSGlWWu}). Under the condition that there is exactly one perturbing atom inside the volume~$V_0$ over which the wavefunction is averaged, one obtains:
\begin{align}\label{eq:fermipotav1}
	\overline{ V_i\psi(\vec{r})}= \frac{4\pi}{V_0}\int V(r)u(r)r\, dr= \nonumber \\ \frac{4\pi}{V_0}  \int \frac{\hbar^2}{2m_e} \frac{d^2u}{dr^2}r\, dr= \frac{2\pi\hbar^2 a}{m_eV_0}\,\overline{\psi}.
\end{align}
The last integral can be solved using integration by parts, under the condition that the wavefunction $\psi$ is differentiable at the position of the perturbing atom~\footnote{This implies that $\left(\frac{du}{dr}r-u\right)_{r=0}=\left(\frac{\partial \psi}{\partial r}r^2\right)_{r=0}=0$.}. If there is no perturbing atom within the volume of interest, the interaction potential vanishes. The general expression of equation~(\ref{eq:fermipotav1}) therefore reads:
\begin{equation}
	\overline{ V_i\psi(\vec{r})}= \frac{2\pi\hbar^2 a}{m_e}\ \frac{\theta(r^*-r)}{V_0}\,\overline{\psi}. 
\end{equation}
Here, $\theta(r^*-r)$ is the Heaviside step function that is zero if $r>r^*$ and is otherwise one.  For the short-range interaction, $r^*$, the function $\theta(r^*-r)/V_0$ converges to the Dirac delta function, so that the interaction potential in the Schr\"odinger equation~(\ref{eq:FermiSGl}) can be written as:
\begin{equation}
	\sum_i V_i(\vec{r})=\int\sum_i V_{\text{pseudo}}(\vec{r}-\vec{R}_i)\left|\psi(\vec{r})\right|^2 \ d\vec{r}, 
\end{equation}
giving the well-known Fermi pseudopotential
\begin{equation} \label{eq:FermiPseudopot}
	V_{\text{pseudo}}(\vec{r})=\frac{2\pi\hbar^2 a}{m_e}\ \delta(\vec{r}).
\end{equation}
This very last step was introduced by Fermi only later in 1936, when treating the scattering of slow neutrons from hydrogen atoms \cite{Fermi1936}. This is the reason why, in the literature, the pseudopotential is commonly associated with nuclear physics. However, the main steps of the derivation were already formulated in 1934~\cite{Fermi1934}. Here, Fermi was only interested in the total effect of a large number of perturbing atoms inside the wavefunction $\psi(\vec{r})$ of the Rydberg electron. By summing up equation~(\ref{eq:fermipotav1}) over all atoms inside a Rydberg atom, he directly obtained the shift, $\Delta$E$(\rho)$, of the Rydberg absorption line, measured by Amaldi and Segr\`e~\cite{Amaldi1934, Amaldi1934a}:
\begin{equation} \label{eq:FermiShift}
	\Delta E(\rho)=\frac{2\pi\hbar^2 a}{m_e}\rho,
\end{equation} 
where $\rho$ is the density of ground state atoms. This equation can be alternatively obtained by first integrating the pseudopotential~(\ref{eq:FermiPseudopot}) over the Rydberg electron probability density, $|\psi(\vec{r})|^2$, to obtain the shift per particle as: 
\begin{equation}\label{eq:singleShift}
\Delta \bar E_\text{single} = \frac{\int_\text{Ryd} V_s(\vecr)d\vecr}{\int_\text{Ryd}d\vecr} = \frac{2 \pi\hbar^2  a}{m_e\mathcal{V}_\text{Ryd}}.
\end{equation}
Subsequently, the shift per particle is integrated over the volume of the Rydberg atom, $\mathcal{V}_\text{Ryd}$, and the relevant \textit{s}-wave scattering potential, $V_s$.  To obtain the total energy shift, $\Delta E$, the mean shift per particle, $\Delta \bar E_\text{single}$, has then to be multiplied by the number of atoms, $N_{\mathcal{V}_\text{Ryd}}$, within a volume as large as the Rydberg atom.  
The interaction potential of ion-perturber scattering events, given by the C$_4/r_{I-P}^4$ potential, leads to a density shift that is three orders of magnitude smaller than the density shift caused by electron-perturber interactions, which are described by the quantum mechanical \textit{s}- and \textit{p}-wave scattering.  Ion-perturber interactions are, therefore, typically neglected at the densities typical in cold atom experiments.  

\begin{figure}
	\centering
		\includegraphics[width=0.85\textwidth]{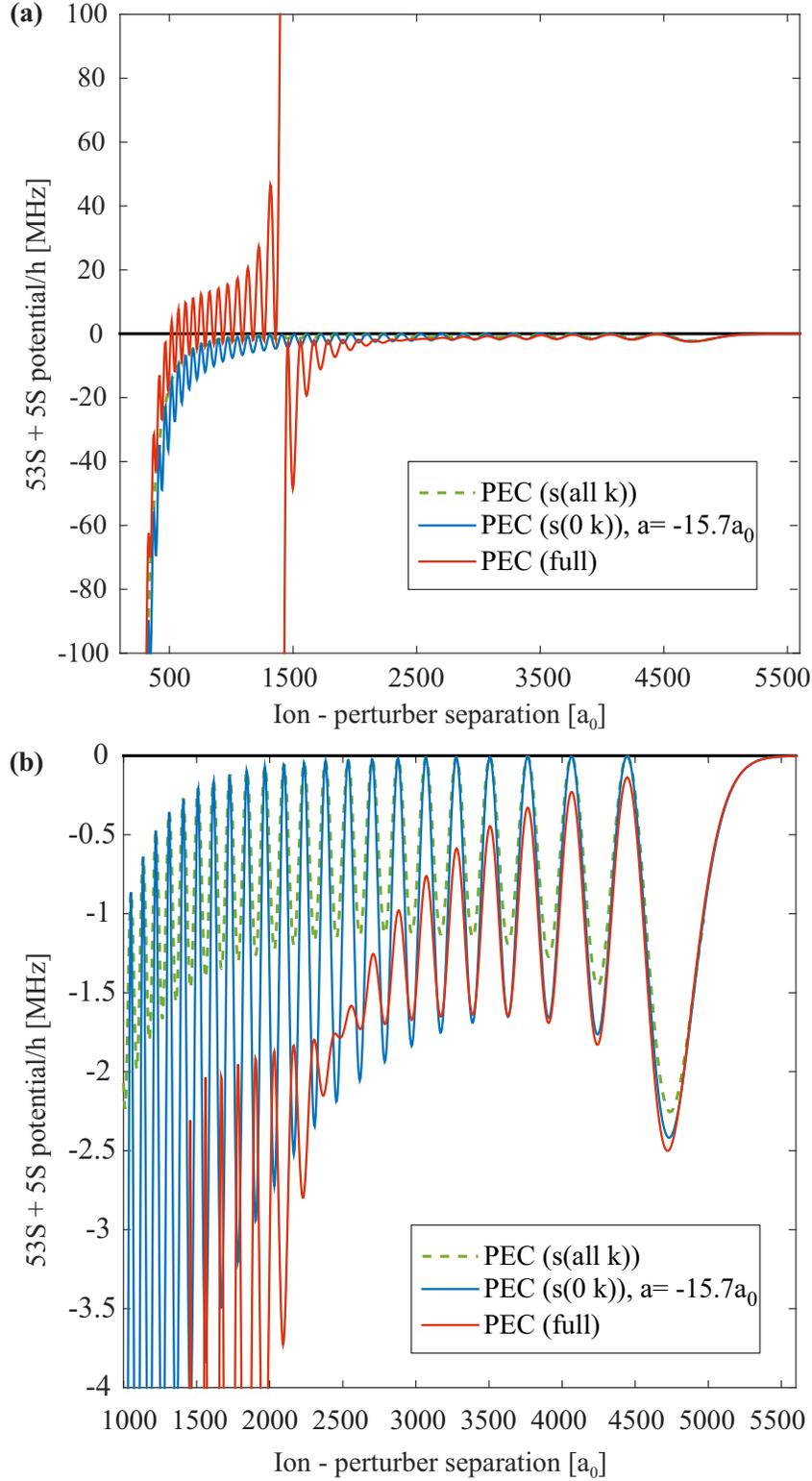}
	\caption{(color online) (a) Adiabatic potential curves for 53\textit{S} + 5\textit{S}, calculated using the different approaches described in the text.  All potential energy curves shown include the ionic polarization potential.  Only the full diagonalization (red), which takes into account \textit{s}- and \textit{p}-wave scattering, produces adiabatic potentials with the so-called butterfly potentials.  (b) A zoom of the potentials shown in (a) to emphasize the differences in the outer potential well. 
	}
	\label{fig:PotComp}
\end{figure}

If there is only one perturber atom inside the Rydberg atom, one would like to calculate the potential energy of a single ground state atom at distance $\vec{R}$ from the center of the Rydberg electron wavefunction $\psi(\vec{r})$ by integrating only over the electron coordinate:
\begin{equation} \label{eq:FermiMolpot}
	V_{\text{scat}}(\vec{R})=\int V_{\text{pseudo}}(\vec{r}-\vec{R})|\psi(\vec{r})|^2\ d\vec{r}=\frac{2\pi\hbar^2 a}{m_e}|\psi(\vec{R})|^2.
\end{equation}
This momentum independent potential energy curve is plotted in Figure~\ref{fig:PotComp} (blue), including the ionic polarization potential, and henceforth referred to as,``PEC(\textit{s}(0 k))".

\subsection{Higher order scattering terms}\label{higherOrder}
The Fermi pseudopotential in its original form is adequate for low-energy electrons, i.e.~for the highly excited Rydberg electron near the classical outer turning point, but the accuracy of the adiabatic potential energy curves can be improved if an energy-dependent scattering length is utilized~\cite{Alekseev1966, Prunele1979, Omont1977}.  The kinetic energy of the Rydberg electron, $E_{\text{kin}}(R)=\hbar^2(k(R))^2/(2m_e)$, can be estimated depending on the radial distance $R$ to the Rydberg core, using a semi-classical approximation \cite{Omont1977}:

\begin{equation} \label{eq:Ekinsemi} 
	E_{\text{kin}}(R)=-\frac{Ryd}{{n^*}^2}+\frac{1}{4\pi\epsilon_0}\frac{e^2}{R},
\end{equation}

where \textit{Ryd} is the Rydberg constant.  
Therefore, as the electron approaches the ionic core, the momentum increases and higher-order partial waves become significant.  For the Rydberg electron-perturber scattering problem, \textit{s}- and \textit{p}-wave contributions suffice, as the interaction for higher order partial waves is suppressed by the centrifugal barrier.  Omont~\cite{Omont1977} developed a method to generalize the Fermi pseudopotential by including contributions of higher partial waves to the electron-atom scattering interaction, whereby the Rydberg electron wave function was expanded in terms of Fourier transformed plane waves.   
The matrix elements of the interaction energy,~$V(\vec{R})$, with the perturbing ground state atom in equation~(\ref{eq:Vpol}) read:
\begin{align} \label{eq:OmontVscat}
	 \left<j\right|V(\vec{R})\left|i\right>= 4\pi a_0^3\sum_{l}(2l+1)\cdot \nonumber\\ R_l\left[P_l\left({\vec{\nabla}'\vec{\nabla}}/{k^2}\right)\psi^*_j(\vec{r}\,')\psi_i(\vec{r})\right]_{\vec{r}=\vec{r}\,'=\vec{R}}, 
\end{align}
where $P_l$ are the Legendre polynomials and the reaction matrix elements $R_l$ are connected to the scattering phase shifts $\eta_l$ by:
\begin{equation}
	R_l = -2Ryd\cdot\frac{\tan{(\eta_l)}}{2a_0k}.
\end{equation}
The first term of the expansion ($l=0$), shown in equation~(\ref{eq:OmontVscat}), leads to the expectation value
\begin{equation}
	V_s(\vec{R})=4\pi a_0^3R_0 |\psi(\vec{R})|^2.
\end{equation}
With the scattering phase shift $\eta_0$ from \cite{Hinckelmann1971}, one obtains 
\begin{equation}
	R_0=\frac{2Ryd}{a_0}\left(\frac{a}{2}+\frac{\hbar^2}{m_ee^2a_0^2}\cdot\frac{\pi}{6}\alpha k\right).
\end{equation}
The first term results in a potential, which is of the form of the molecular potential~(\ref{eq:FermiMolpot}) as calculated from Fermi's pseudopotential~(\ref{eq:FermiPseudopot}):
\begin{equation} \label{eq:swavepot}
	V_s(\vec{R})=\frac{2\pi\hbar^2}{m_e}a(k) |\psi(\vec{R})|^2.
\end{equation}
It describes \textit{s}-wave scattering with a momentum-dependent \textit{s}-wave scattering length:
\begin{equation} \label{eq:akedp}
	a(k)=a+\frac{\hbar^2}{m_ee^2a_0^2}\cdot\frac{\pi}{3}\alpha k +O(k^2).
\end{equation} 
Using this position dependent scattering length an altered adiabatic potential energy curve is plotted in Figure~\ref{fig:PotComp} (green, dashed), and henceforth referred to as ``PEC(\textit{s}(all k))".    

For Rydberg \textit{S}-states, the PEC(\textit{s}(all k)) supports multiple bound states of one or more neutral atom perturbers in the outermost potential well, shown in Figure~\ref{fig:PotComp} at $\approx\,4700$\rbohr.  This is the basis of ultralong-range Rydberg molecules first predicted in 2000~\cite{Greene2000} and first observed in 2009~\cite{Bendkowsky2009}.  The PEC(\textit{s}(all k)) given by equation~(\ref{eq:FermiMolpot}) and shown in Figure~\ref{fig:PotComp} (green) represent the contribution of the electron momentum~$\hbar k$ to the momentum dependent triplet \textit{s}-wave scattering length (equation~(\ref{eq:akedp})).  The PEC(\textit{s}(all k)), furthermore, predicted the existence of trilobite-like polar homonuclear diatomic molecules~\cite{Greene2000} for triplet \textit{s}-wave scattering, which were observed in~\cite{Li2011}.  The PEC(\textit{s}(all k)) are a good approximation for calculating the binding energies of Rydberg molecules in the outer two to three potential wells, whereby the bound state is localized at large internuclear distances, and \textit{s}-wave scattering dominates.  However, at internuclear distances less than 2000\rbohr,~\textit{p}-wave scattering dominates, as is described next in section~\ref{Resonance}, and the validity of PEC(\textit{s}(all k)) fails. 

The accuracy of the potential energy curves analyzed with the Omont-Fermi approach depend on precise \textit{s}- and \textit{p}-wave scattering phase shifts for the triplet and singlet configurations.  Values of the electron-alkali atom scattering lengths,~$a$, were obtained from Dirac R-matrix calculations~\cite{Bahrim2001}. Depending on the species of the perturbing atom and the relative orientation of the electron spin and atomic spin, the value can be positive or negative as pointed out by Fermi~\cite{Fermi1934} and verified by Reinsberg~\cite{Reinsberg1934}. This could explain the blue and red shifts observed in~\cite{Amaldi1934,Amaldi1934a,Fermi1934}. Alkali atoms have a large electron affinity leading to typically negative scattering lengths.  For $^{87}$Rb the triplet scattering length, relevant for the experiments described here, is $a_{\uparrow\uparrow}=-15.7\,a_0$~\cite{Boettcher2015}.  The singlet scattering length is significantly smaller; it was recently measured to be $a_{\uparrow\downarrow}=-0.2\,a_0$~\cite{Boettcher2015}.  The triplet and singlet scattering lengths were first theoretically derived in~\cite{Bahrim2001} to be $a_{\uparrow\uparrow}=-16.1\,a_0$ and $a_{\uparrow\downarrow}=0.627\,a_0$. 

To the same order in the electron momentum,~$\hbar k$, the second term in the expansion leads to a \textit{p}-wave contribution as outlined by Omont~\cite{Omont1977}:
\begin{align}
V_{\rm{Ryd-perturb}}( \vec{r},\vec{R} ) = V_{\rm{Fermi}}( \vec{r},\vec{R} )- \nonumber \\
\frac{6\pi \tan{\left(\delta^{\textit{p}}[k(R)]\right)}}{k(R)^{3}}\delta^{(3)}(\vec{r}-\vec{R})
\overleftarrow{\nabla}_{r}\cdot \overrightarrow{\nabla}_{r},
\label{eq:Omont}
\end{align} 
where $\delta^{\textit{p}}[k(R)]$ stands for the triplet \textit{p}-wave scattering phase shift of an electron-perturber collision.  The arrow indicates that this should be read as an operator on the wavefunction, acting in the indicated direction.  Quasi-degenerate perturbation theory has been applied to calculate the Born-Oppenheimer PECs for the electron-atom interaction, henceforth referred to as ``PEC(full)", which are shown in Figure~\ref{fig:PotComp}(red). To calculate PEC(full) a hydrogen-like basis for high angular momentum states, (l~$\geq3$), has been used, whereas the Whittaker functions~\cite{Abramowitz1974} have been utilized for low angular momentum states, (l~$\leq3$), thus taking into account the quantum defects. For each diagonalization eight different \textit{n}-manifolds have been included as well as their respective angular momentum states; in particular, two of them are above the target state and six are below, taking into account the correct energy ordering of the states due to the quantum defects.  For an extended description of the calculation see~\cite{Hamilton2002}.  

\subsection{\textit{p}-wave shape resonance}\label{Resonance}  

In some systems there is a resonant bound state behind the centrifugal barrier.  Indeed, in alkali atom systems there is a \textit{p}-wave shape resonance, which depends on the shape of the centrifugal barrier.  The resonance leads to deep potentials, which span from one hydrogenic manifold, \textit{n}, to the next hydrogenic manifold, (\textit{n}-1), as shown in Figure~\ref{fig:PotComp}, predominantly due to the scattering phase shift, $\delta^{\textit{p}}$, of $\pi$.  The \textit{p}-wave shape resonance deviates from the typical $n^{-6}$ scaling of the binding energy found for the outer potential wells of the electron-perturber adiabatic potentials, as shown in Figure~\ref{fig:PotComp} and shown in~\cite{Gaj2014}.  The depth of the \textit{p}-wave potential, extending from the hydrogenic manifold, \textit{n}, to the well of the potential, scales rather as $n^{-3}$.  The \textit{p}-wave potential is oscillatory and each local minimum along the potential curve can support multiple bound states, depending on the \textit{n} investigated and the reduced mass of the Rydberg-perturber system.  The \textit{p}-wave potential energy curves can be viewed as a linear combination of electronic states that maximizes the derivative of the wave function parallel or perpendicular to the internuclear axis~\cite{Chibisov2002, Hamilton2002}.  This resulting electron wave function has a butterfly-like structure giving rise to the naming convention of the \textit{p}-wave scattered states to be noted as "butterfly states"~\cite{Hamilton2002}.  In the case of \rubidium, the crossing of the \textit{p}-wave potential with the \textit{nS} states, which are depressed in energy from the hydrogenic manifold due to quantum defects, occurs at $\approx1400\rbohr$ for $n=40$ and saturates  for $n>100$ at interatomic spacings of $\approx\,1800\rbohr$.  Henceforth the crossing of the \textit{p}-wave resonance potential with the \RydGS, \RydGD, and \RydGP~potentials will be represented with the symbols $C_S$, $C_D$, and $C_P$ respectively, and generically with the symbol $C_l$.  The slope of the crossing, $C_S$, scales as \textit{n}$^{-4}$ for principal quantum numbers 40\,$<$\,\textit{n}\,$<$\,111.   

The crossings, $C_l$, lead to large \textit{n}-dependent deviations in the broadening of the Rydberg spectra.  This was observed in thermal cell experiments where the line shift and line broadening were measured from $n=15$ to $n=40$~\cite{Thompson1987}.  Modeling the dependence of the line shape broadening versus principal quantum number, led to the prediction of a \textit{p}-wave shape resonance~\cite{Fabrikant1986}.  Subsequent \textit{ab initio}~calculations confirmed that Rubidium has a shape resonance in the \textsuperscript{3}\textit{P}\textsuperscript{0} symmetry for the electron-Rb(5\textit{S}) scattering at $23\,\text{meV}$~\cite{Bahrim2001}.  The scattering potential is affected in the range of approximately $10\,\text{meV}$ around the resonance~\cite{Bahrim2000,Hamilton2002,Greene2006}.  Evidence of the \textit{p}-wave shape resonance was recently observed and modeled via Rydberg spectroscopy in a BEC~\cite{Schlagmueller2015}.  The \textit{p}-wave shape resonance also plays a role in changing the binding energy of long-range Rydberg molecules.  It has been shown that excited states become bound due to internal quantum reflection at the \textit{p}-wave shape resonance~\cite{Bendkowsky2010}.  It has also been shown that the binding energy of ground state Rydberg molecules, bound in shallow potentials resulting from hyperfine mixing of triplet and singlet potentials, is sensitive to the triplet \textit{p}-wave scattering phase shifts at low energies~\cite{Boettcher2015}. 

Modifications of the PEC(full) presented here have been calculated.  Methods beyond the Omont-Fermi method have been developed to calculate the potential energy curves of the Rydberg-perturber system to small ion-neutral separations.  The potential energy curves of the \textit{p}-wave resonance have also been calculated using the Green's function method~\cite{Chibisov2002, Schlagmueller2015} and good agreement is found with PEC(full) shown in Figure~\ref{fig:PotComp}~(blue).  Work by Khuskivadze et~al.~\cite{Khuskivadze2002} extended the calculations of the potential energy curves for the \textit{p}-wave resonance by including spin-orbit coupling, and an unrestricted electron-perturber interaction radius.
Even within the Omont-Fermi model, calculations by Anderson et~al.~\cite{Anderson2014A} have demonstrated modifications of the potential energy curves resulting from inclusion of the hyperfine splitting of the alkali atom perturber.  Hyperfine splitting of the perturber atom leads to a second-order mixing of singlet and triplet scattering terms and a new set of shallow potential energy curves that support bound states.  This prediction of Rydberg molecules for these shallow potentials have been demonstrated experimentally for Cs~\cite{Sassmannhausen2015} and Rb~\cite{Boettcher2015}.  These shallow PECs were most certainly relevant in the experiments of Thompson et al.~\cite{Thompson1987} performed in a thermal cell with a non-spin-polarized atomic sample.  

\subsection{Beyond Fermi's mean-field density shift}\label{densityShift}
 \begin{figure}
	\centering
		\includegraphics[width=0.9\textwidth]{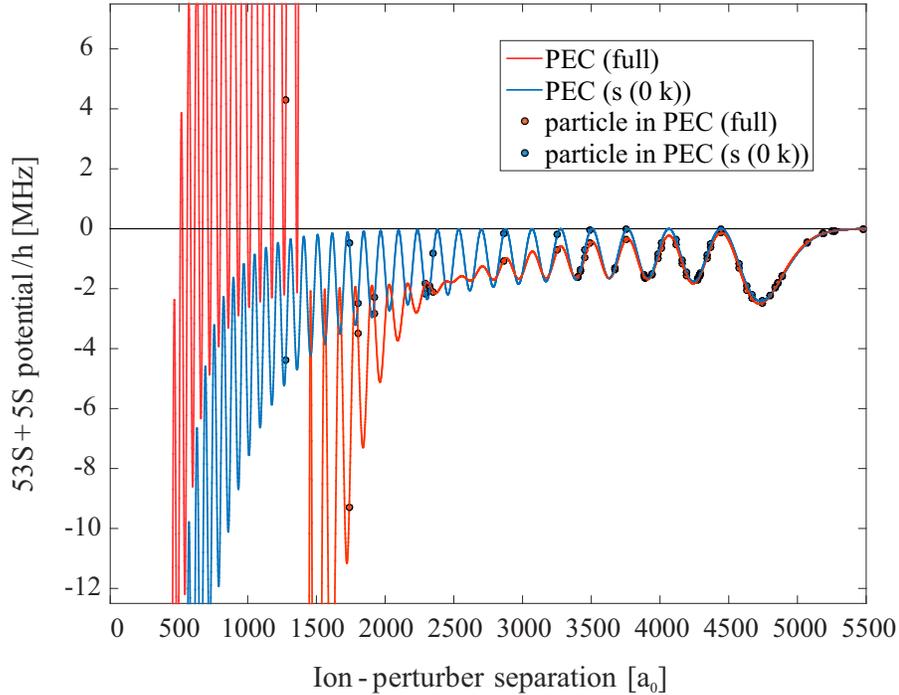}
	\caption{(color online) Potential energy curves for 53\textit{S}\,+\,5\textit{S}, plotted for \textit{s}- and \textit{p}-wave scattering and \textit{s}-wave scattering only, both including the ionic core polarization potential. Discrete particles are randomly placed in the potential according to the peak density of the BEC of \density{5.2}{14}, which has a mean interparticle distance of 1300\rbohr. The available energy range for the background perturber atoms is increased for  \textit{s}- and \textit{p}-wave scattering (PEC (full)) due to the crossing, $C_S$. For this particular configuration, with 37 particles within the classical Rydberg orbit, the total density shift would result in {-58.2}\MHz~for \textit{s}- and \textit{p}-wave scattering (blue) and {-51.2}\MHz~for \textit{s}-wave scattering only (red).
	}
	\label{fig:PEC53SShiftExample}
\end{figure}
The \textit{p}-wave shape resonance leads not only to \textit{n}-dependent line broadening~\cite{Stoicheff1980, Thompson1987, Fabrikant1986, Schlagmueller2015, Borodin1991}, but also to deviations from the \textit{n}-independent mean-field density shift, calculated using Fermi's approach as outlined in section~\ref{Pseudopotential}.  The origin of this deviation from Fermi's \textit{n}-independent mean-field density shift is because the slope of the crossing of the \textit{p}-wave resonance potential with the \RydGS~potential, (noted here as $C_S$), does not scale inversely with the Rydberg orbital volume, i.e.~as \textit{n}$^{-6}$, but rather as \textit{n}$^{-4}$ (see section~\ref{Resonance}).  The ion-perturber separation at which the crossing, $C_S$, occurs, is also \textit{n}-dependent as discussed in section~\ref{Resonance}. 
The \textit{p}-wave potentials lead, therefore, not only to deviations from Fermi's \textit{n}-independent mean-field density shift, but also a density shift that is dependent on the density inside the Rydberg orbit.  At high densities the nearest neighbor particle spacing approaches values close to values of $C_l$, making the deviations from Fermi's mean-field density shift, due to the \textit{p}-wave shape resonance, more significant at higher densities.   We consider in this review only the change of the density shift pertaining to Rb \textit{nS} states as a proof of principal concept.  

In Figure~\ref{fig:PEC53SShiftExample} classical hard-sphere background perturber atoms for a nearest neighbor distribution, given by a peak density of \density{5.2}{14}, are plotted with respect to PEC(\textit{s}(0 k)) and PEC(full).  With such a representation it is straightforward to denote the total energy shift, $\delta_{i}$, of each possible Rydberg excitation within the atomic sample located at $r_i$,~\cite{Schlagmueller2015} as,
\begin{equation} \label{eq:energshift}
	\delta_i=\sum_{j\neq i} V\left(\left|\vec{R_j}-\vec{r_i}\right|\right), 
\end{equation} 
where $\vec{R_j}$ is the location of the background perturber atom.  This model was employed in~\cite{Schlagmueller2015}.  Using this model the \textit{n}-dependent line broadening was simulated utilizing the PEC(\textit{s}(all k)) and the PEC(full).  Good agreement with the line broadening of Rydberg spectra taken in a BEC was found only for the case of PEC(full), demonstrating the significance of the \textit{p}-wave shape resonance for alkali-atom Rydberg spectroscopy, when there is a high probability for ion-perturber separations less than $2000$\rbohr.  

\begin{figure}[t!]
	\includegraphics[width = 1\columnwidth]{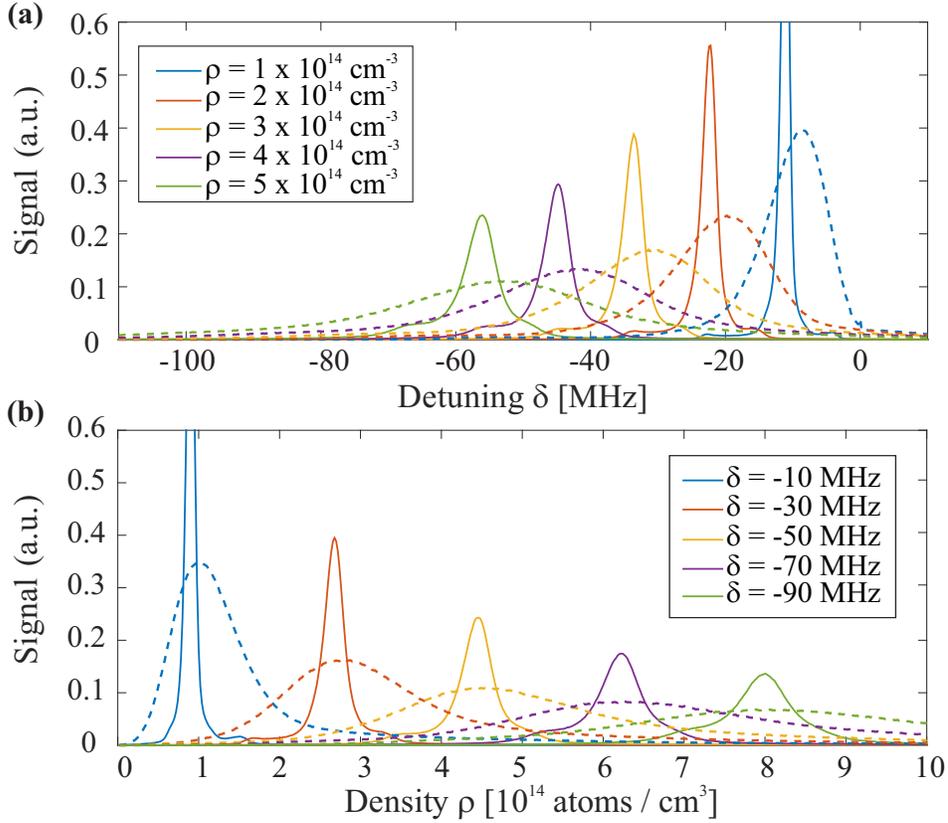}
	\caption{(color online) (a) Simulated Rydberg spectra for constant BEC densities of a 53\textit{S} state (dashed line) and a 111\textit{S} state (solid line), with the 53\textit{S} signal enhanced by a factor of 2.5 for better visibility. The 111\textit{S} state shows a narrower line compared to the broad signal for lower \textit{n}. (b) The addressed density distribution for a constant laser detuning as the inverse of the parameters of (a) shows for high \textit{n}, that a narrower bandwidth of densities can be addressed by the laser detuning. 
}
	\label{fig:Configurations_vsn}
\end{figure}

Measuring Rydberg atom spectra in dense atomic samples is a measurement of the probabilities of background atom perturber configurations possible within the Rydberg orbit, for a given background atom density distribution.  To simulate the \textit{n}-dependent spectroscopic line shifts, densities are sampled with a resolution of \density{0.1}{14} from $10\textsuperscript{13}\,\text{cm}\textsuperscript{-3}$ to $10\textsuperscript{15}\,\text{cm}\textsuperscript{-3}$, and at each density five million atomic position configurations are calculated.  The density is homogeneous across the Rydberg orbit in the simulations presented here, as opposed to in~\cite{Schlagmueller2015}, where an inhomogeneous density across the Rydberg orbit was examined.  The simulations comparing a homogeneous versus inhomogeneous background density produce indistinguishable line shape results to those presented in~\cite{Schlagmueller2015}.  We calculate the probability of the resulting density shift, or equivalently, the resulting required photoassociation laser frequency, due to the possible configurations within the Rydberg orbit at a particular density.  Once the probabilities of possible energy shifts are calculated over the range of densities, the results are then mapped into signal versus density for chosen detunings.  The results of both of these calculations are shown in Figure~\ref{fig:Configurations_vsn} for 53\textit{S} and 111\textit{S}.  Figure~\ref{fig:Configurations_vsn}(b) demonstrates the bandwidth of densities over which the Rydberg atom excitation is created for a given laser detuning.  For 111\textit{S} the bandwidth of densities where the Rydberg excitation is created, is narrower than for 53\textit{S}.  For 53\textit{S} the large energy shift deviations caused by $C_S$, lead to many probable configurations of background atoms, i.e.~various background densities, that lead to the same total energy shift, i.e.~chosen laser detuning.  This simulated result demonstrates that for a chosen laser frequency the Rydberg excitation is well localized to a density shell within the atom cloud for high principal quantum numbers, whereas for low principal quantum numbers, the Rydberg atom excitation is not as well localized.  Experimental results demonstrating the localization of a Rydberg excitation for \textit{n}\,$=$\,90 are shown in Figure~\ref{fig:offset_scan}.    

\begin{figure}
		\includegraphics[width=0.85\textwidth]{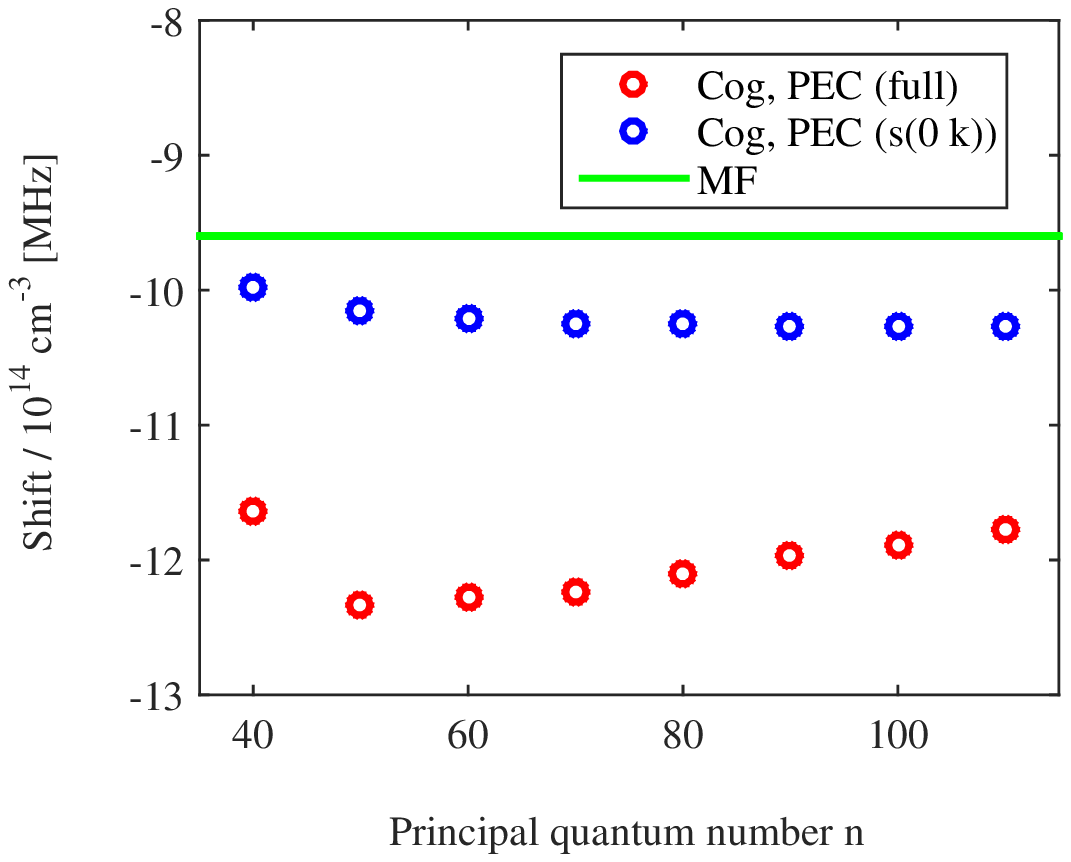}
	\caption{(color online) Plotted are the center of gravities (Cog) of the spectra simulated using the point-particle perturber model for PEC(full) (red) and PEC(\textit{s}(0 k)) (blue) (See Figure~\ref{fig:PEC53SShiftExample}) versus $n$ for a fixed density, $\rho$, of \density{5}{14}. For example the spectra for $\rho=$\density{5.2}{14} (green) shown in figure~\ref{fig:Configurations_vsn}(a) for $n$=53 and 111 correspond approximately to the data points for $n$=50 and 110 plotted. The straight line (green) is the density shift given by Fermi's mean-field model using a triplet scattering length of -15.7\rbohr.  Using Fermi's model there is no $n$-dependence of the mean-field density shift, but the point-particle perturber model predicts a reduction of the density shift for low-\textit{n} due to Poissonian fluctuations of the atom number and deviations of the wavefunction from the $n^{-6}$ scaling.  
	}
	\label{fig:DensityShift}
\end{figure}

We use the simulated signal versus detuning spectra shown in Figure~\ref{fig:Configurations_vsn}(a) to calculate the expected density shift versus principal quantum number as shown in Figure~\ref{fig:DensityShift}.  The spectral center of gravity (Cog) was calculated to determine the density shift given by the simulated spectra.  The spectra for all \textit{n} at a homogeneous density of \density{5}{14} were used. Within this model the density shift is not absolute for all densities.  The \textit{n}-dependent density shift for PEC(\textit{s}(0 k)) shown in Figure~\ref{fig:DensityShift}, is due to Poissonian fluctuations of the atom number at low $n$ and is exacerbated by the deviations of the wavefunction from the $n^{-6}$ scaling.      
This can be understood by considering equation~\ref{eq:singleShift} for the case of small $N_{\mathcal{V}_\text{Ryd}}$.  The contribution of each perturber atom, within the Rydberg orbit, to the total density shift becomes increasingly more significant in the limit that $N_{\mathcal{V}_\text{Ryd}}$ goes to zero.  For example this is the case for low \textit{n} for the range of densities (\density{1-10}{14}), examined here.  At a density of \density{5.2}{14}, which is a high density compared to most cold atom experiments, there are only $5.5\pm2.5$ atoms within the volume of the Rydberg orbit for \textit{n}\,$=$\,40.  The deviations of the density shift for low $N_{\mathcal{V}_\text{Ryd}}$ is most pronounced in the case of PEC(full), due to the $n^{-4}$ scaling of $C_l$.  In the case of low $N_{\mathcal{V}_\text{Ryd}}$, perturber atoms excited near $C_S$ introduce large energy shift contributions that lead to a reduced density shift, as shown in Figure~\ref{fig:DensityShift}. This dependence of the line shift with \textit{n} found with the point-particle perturber model presented here, is in agreement with the findings of~\cite{Fabrikant1986, Borodin1991}.  The slope of line broadening versus \textit{n}, shown in~\cite{Thompson1987}, fit best by taking the \textit{p}-wave shape resonance into account, leading to the first prediction of a shape resonance for electron-alkali atom scattering.  

The deviations of the simulated density shifts for 40$\,<$$\,n\,$$<\,$110 using PEC(\textit{s}(0 k)) (blue) from the mean-field result (green), as shown in Figure~\ref{fig:DensityShift}, is due to the inclusion of the polarization potential.  If the polarization potential is neglected, the simulated density shifts for $n\,$$>\,$50 have the same values as the $n$-independent mean-field density shift.   

There are limitations of the point-particle perturber model, presented here and in~\cite{Schlagmueller2015}.  As shown in Figure 3 of~\cite{Schlagmueller2015} the modeled spectra do not agree well for~\textit{n}$>53$.  Because of this discrepancy, we simulate density shifts that are larger than what we measure experimentally, as will be discussed further in section~\ref{Comparison}.  One reason for this discrepancy of the simulated spectral line shifts could be that the point-particle perturber model is a purely classical consideration of the background atoms.  Despite not solving for potential bound states of PEC(full), the overall resulting density shift should be captured by the point-particle perturber model.  This discrepancy suggests that the non-local wavefunction of the ground state atoms might need to be considered. Perhaps refinements of the potential energy curves are also required to find better agreement between the simulated density shift using PEC(full) and the experimental measurements.  Examples of the extended theories for calculating potential energy curves are discussed in section~\ref{Resonance}.  A possible limitation of the point-particle perturber model presented here, is that it is effectively a two-body model, although the resulting shift of many particles are considered.  This is because only pairwise PECs are considered, and any resulting alteration of the Rydberg electron density due to a perturber excited within $C_l$, is neglected.    

Using classical hard-spheres, with zero temperature, for simulating the effect of background gas perturbers on the Rydberg spectrum taken in a BEC, is justified because of the timescale of the experiment.  The chemical potential of a typical BEC, used in our experiments, is on the order of 1\kHz, leading to BEC response times on the order of milliseconds.  Therefore, no collective response of the BEC is expected on the microsecond timescale of the excitation pulses used in the experiment, as will be discussed in section~\ref{Measurement}.  Assuming zero temperature also neglects any resulting bunching expected in a thermal cloud. 

Schimdt et al.~\cite{Schmidt2015} study the line broadening of a single Rydberg atom in a high density medium from a many-body perspective. In particular, all the different orders of correlations among particles are included in the model, which gives overall agreement with the line shape of the polyatomic Rydberg molecules. The agreement was excellent for low detunings but poorer for higher detunings, which might indicate a need for a more precise characterization of the physics at the few-body level. Indeed, the model of Schmidt et al.~begins from few-body information: namely the two-body bound states, which were included by a fitting procedure to match the experimental values~\cite{Schmidt2015}. While this theory is based on a many-body treatment, it is not explicitly compared with results of the more common two-body thermally-averaged formulations of line-broadening theory, which leaves open the question of how essential it is to implement a full many-body description in this system. The polyatomic nature of Rydberg aggregates has also been recently considered in~\cite{Eiles2016}. 

\section{High density Rydberg spectroscopy}\label{Spectroscopy}
\subsection{Measurement technique}\label{Measurement}
\begin{figure}[th!] 
	\includegraphics[width=0.9\textwidth]{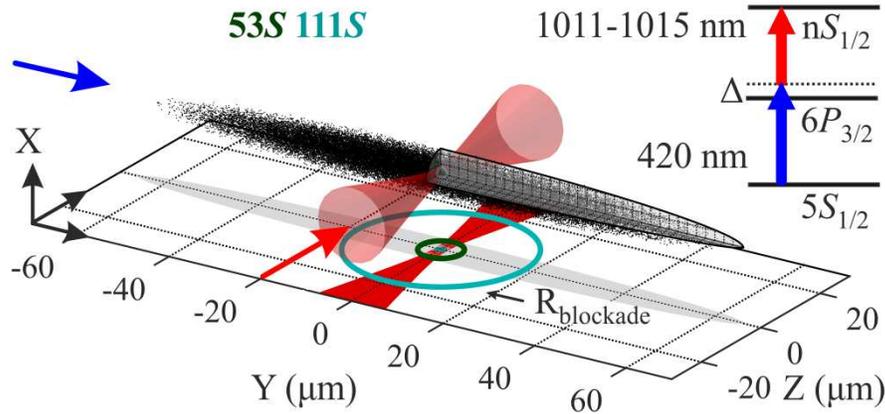}
	\caption{(Color online) Simplified schematic of a localized Rydberg excitation in our BEC drawn to scale. A focused infrared laser along the z-axis excites Rydberg atoms and determines the excitation volume inside the BEC together with a \unit{420}{nm} collimated beam along the y-axis.  The extent of the classical Rydberg electron orbital for the highest Rydberg state under investigation (111\textit{S}) is shown as a tiny solid sphere (cyan) at the very center (filled circle on projection).  The extent of the Rydberg blockade, assuming a 5\MHz~excitation bandwidth, is shown for 111\textit{S} (cyan) and 53\textit{S} (green).  A simplified level scheme of our two-photon excitation is illustrated in the upper right corner.}
	\label{fig:excitation_scheme}
\end{figure}

We perform Rydberg spectroscopy in a nearly pure BEC produced in a QUIC trap~\cite{Esslinger1998}.  The BEC has approximately \becatoms~\rubidium~atoms in the magnetically trapped ground state $\ket{\rm{5}\textit{S}_{\rm{1/2}}\rm{,\,}\textit{F}\rm{\,=\,2,\,}\textit{m}_{\textit{F}}\rm{\,=\,2}}$. The trapping frequencies are $\omega_{r}$\,=\,2$\pi\times$200\Hz~in the radial and $\omega_{ax}$\,=\,2$\pi\times$15\Hz~in the axial direction corresponding to Thomas-Fermi radii of 5.0\um~by 66\um~(see Figure~\ref{fig:excitation_scheme}).  The atom number and trap frequencies give rise to a peak density of the BEC of \density{5.5}{14}, just after the RF-field, used for evaporative cooling, is turned off.  The Rydberg excitation lasers are turned on 1\us~of the RF-field being turned off.  For the Rydberg excitation we apply a two-photon excitation scheme, where we couple the ground state to the $\ket{\textit{nS}_{\rm{1/2}}\rm{,\,}\textit{m}_{\textit{S}}\rm{\,=\,1/2}}$ state for principal quantum numbers \textit{n} from 53 to 111, via the intermediate state 6\textit{P}$_{\rm{3/2}}$, as sketched in Figure~\ref{fig:excitation_scheme}. For the lower transition we illuminate the cloud along the axial direction with pulsed 420\nm~light, 2\mm~in waist, to ensure a uniform Rabi frequency across the entire cloud. We use an intermediate state detuning of $\Delta$\,=\,80\MHz~$\approx$\,56\,$\Gamma$ in order to keep absorption and heating of the BEC, due to scattering of the 420\nm~light, low. The upper transition to the Rydberg state is driven by an infrared laser with an \textit{n}-dependent wavelength between 1011\nm~and 1015\nm, focused down to a (2.1$\pm$0.3)\um~waist, and is applied parallel to the z-axis, as shown in Figure~\ref{fig:excitation_scheme}.  The 420\nm~excitation light is circularly polarized ($\sigma+$), and the infrared light is set to have linear polarization along the x-axis perpendicular to the magnetic field in the y-direction. We pulse both excitation lasers simultaneously for 2\us, with a repetition rate of 2\kHz~to create Rydberg atoms, which we subsequently electric field ionize with a field strength of 156\Volt$/$\cm~within 400\,ns~of the light fields turn-off.

It is crucial that particular attention is paid to the ionization field turn-on time and the ionization field strength when studying Rydberg spectroscopy in dense background gases.  As discussed in subsequent work, $l$-changing collisions occur due to collisions of the Rydberg atom and the neutral atom perturber near the core, on timescales of less than 1\us~for $n<90$ at densities greater than \density{1}{14}~\cite{Schlagmueller2016PRX}.  Rydberg atoms that have undergone an $l$-changing collision ionize at an ionization field up to three times higher than the ionization threshold field of the original \textit{S}-state~\cite{Gallagher}.  Therefore, if the ionization field is set to the threshold ionization field, the spectral signal obtained from the highest densities that underwent a state-changing collision on a fast timescale, will be suppressed. With good control of the excitation laser pulse timing, and the ionization field timing, the ionization threshold voltage could be used as a feature to alter the spectral line shape versus the collisional lifetime.  Another important experimental detail critical for measuring the line shift of the spectra is that signal saturation effects must be avoided.  The excitation laser power was kept low, resulting in an average ion count of 0.4, such that saturation of the signal due to the Rydberg blockade and the MCP detector were kept low.  We estimate the detection efficiency of the MCP in our experiment to be 70\%.  It is also important that a fairly large scan range must is used in order to measure all spectroscopic signal at large red and blue detunings, otherwise the line shift measurement via, for example, the center of gravity of the spectrum, could be distorted.  Simulations of the spectroscopic line shape~\cite{Schlagmueller2015} revealed that the spectroscopic data taken in the BEC and presented here did not indeed cover a large enough detuning range, as will be discussed further in section~\ref{effectiveDensity}.  

\begin{figure}[t!]
	\includegraphics[width=0.9\textwidth]{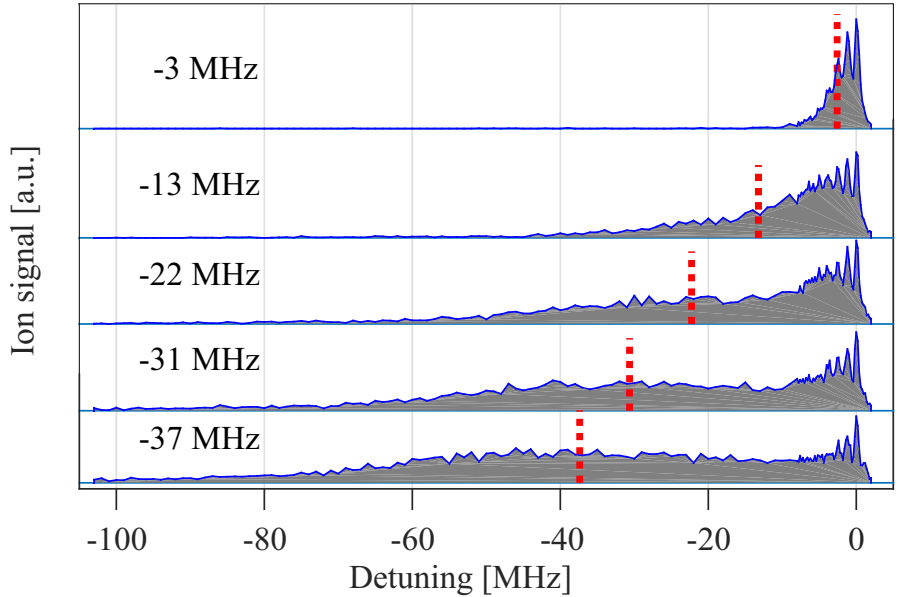}
	\caption{(color online) Spectra of the 53\textit{S} state for the same initial atom cloud, but different mean densities. The ion signal scaling is the same for all spectra. The bottom trace is a spectrum in a nearly pure BEC, for the first 50 excitation pulses, and is the same spectrum as displayed in figure 2 of~\cite{Schlagmueller2015}.  The top spectrum is signal from the last 50 out of 1000 excitation pulses when the atomic cloud is a thermal cloud.  The dotted lines indicate the center of gravity (Cog) for each mean density, and the values of the Cog are noted to the left of the spectra.}
	\label{fig:53SSpectra}
\end{figure}

For the spectra shown in Figure~\ref{fig:53SSpectra}, each spectral point is taken in a new BEC, with the detuning of the infrared laser changed between successive clouds. The two-photon, single atom Rabi frequency of 250\kHz~is kept constant for all spectra taken for \textit{n} from 53 to 111, by increasing the power of the infrared laser for higher Rydberg states. For the highest applied power, the trap depth of the time-averaged optical potential for this laser is about 100\nK, which is less than the starting temperature of our BEC of $\approx$\,300\nK.  Due to low Rabi frequencies and strong van der Waals interactions between Rydberg atoms~\cite{Saffman2010}, the probability of creating multiple Rydberg excitations within the excitation volume, defined by the infrared laser, is very low. 

\subsection{Effective density measured via spectra}\label{effectiveDensity}
We analyze the density shift for spectra taken for \textit{n} from 53 to 111 in the BEC.  We apply 1000 excitation and ionization cycles in a single BEC within 500\,ms.  During the excitation pulses the BEC decays into a purely thermal cloud due to atom loss, allowing us to sample a range of densities and temperatures.  Over the 500\,ms~timescale of excitation pulses, we lose about half of the total atoms mainly due to ground-state three-body collisions~\cite{Soeding1999}.  Due to off-resonant scattering from the $6$\textit{P}$_{3/2}$ state, caused by the 420~\nm~light pulses, we lose less than 500 atoms from the cloud per excitation pulse for \textit{T}$<$\textit{T}$_{\rm{c}}$, where \textit{T}$_{\rm{c}}$ is the critical temperature.   The off-resonant scattering from the $6$\textit{P}$_{3/2}$ state also causes some heating of the cloud.  After $1000$ pulses the final thermal cloud has a temperature of $600$\nK~with the $420$\nm~light blocked, and $770$\nK~with the $420$\nm~light unblocked.  For the chosen excitation and detection parameters, additional atom loss due to Rydberg-caused phonon excitations in the BEC~\cite{Balewski2013} is minimal.

Figure~\ref{fig:53SSpectra} shows the evolution of the 53\textit{S} state spectra as a function of an effective density (defined below), as the BEC and eventually the thermal cloud, decrease in density. We extract these spectra by splitting the 1000 excitation pulses, taken in a single cloud, into blocks of 50.  We average over fifty pulses to increase the signal-to-noise ratio. The density over 50 excitation pulses varies by $\approx$\,10\% for \textit{T}$<$\textit{T}$_{\text{c}}$, and less than 2\% for \textit{T}$>$\textit{T}$_{\rm{c}}$.  Each point in all shown spectra represents 500 measurements, taken in 10 different clouds, to improve the signal-to-noise ratio.  As the effective density in the cloud decreases, the spectra congruously narrow and shift towards lower red detunings, as expected from the density shift.  The peaks $\gtrsim-10$\MHz~ in the spectra shown in Figure~\ref{fig:53SSpectra} are attributed to Rydberg excitations in the surrounding thermal cloud, which has a peak density of \density{1.8}{13}. These sharp peaks are due to Rydberg-ground state molecules~\cite{Bendkowsky2009} (see section~\ref{higherOrder}) and indicate the spectral resolution that we are sensitive to. Our spectral resolution is limited to $\approx$ 450\kHz~by the Fourier width of the excitation pulses. 

The Rydberg spectra averaged over the first $50$ excitation pulses represent Rydberg spectra for the starting conditions of the BEC, i.e.~the densest and coldest conditions of the experiment.  A spectrum representing these conditions is shown for \textit{n}\,=\,53 in the bottom panel of Figure~\ref{fig:53SSpectra}.  Spectra for the initial conditions of the BEC for \textit{n}\,=\,53, 71, 90, and 111 are shown in Figure 3 of~\cite{Schlagmueller2015} and are discussed in detail there.  There is an \textit{n}-dependent broadening as observed in previous experiments in thermal cells~\cite{Thompson1987} and in the BEC~\cite{Schlagmueller2015}.   Therefore, in order to measure a mean density shift, $\Delta E(\rho)$~(see equation~\ref{eq:FermiShift}), versus \textit{n}, that is not sensitive to broadening, it is more appropriate to compare the center of gravities (Cog) of the spectra.  It is of great interest to compare these measured Cogs to the expected mean-field density shift, using Fermi's model outlined in section~\ref{Pseudopotential} and the point-particle model using the appropriate PECs as described in section~\ref{densityShift}.  In order to make this comparison, it is necessary to experimentally measure the effective densities of our excitation region, as described in the next section, section~\ref{measuredDensity}.

\subsection{Effective density measured via time of flight images and beam profiling}\label{measuredDensity}
We obtain the effective density of our system by calculating the overlap integral of the excitation beam with a Gaussian intensity profile of $I(\vec{r})$, with the bimodal density distribution, $\rho$. The relative atom numbers of condensed and thermal atoms were calibrated by using a bimodal distribution to analyze off-resonant and on-resonant time-of-flight absorption images, respectively.  The temperature was calibrated using the thermal fraction measured from on-resonant time-of-flight absorption images.  The atom number and temperature of the atom cloud was measured at intervals of 50 excitation pulses, corresponding to the intervals used for analyzing the spectra.  Extra measures were taken to ensure that the heating of the coils used for the QUIC trap had a constant load, independent of the time interval that the excitation pulses were applied.  This was necessary to achieve a temperature and density calibration consistent with the conditions of the atom cloud at the time of excitation.  We use the calibration measurements to calculate a position dependent effective excitation probability, \textit{P}$_{\rm{exc}}$ as,
\begin{equation}
	P_{\rm{exc}(\vec{r})} \propto \frac{ \rho~I(\vec{r})}{1 + 4 (\frac{\delta(\vec{r})}{\gamma})^2}     
	\label{eq:scattering_rate}
\end{equation}
where $\delta(\vec{r})$ is the detuning from the two-photon Rydberg resonance due to the laser detuning and the density shift, $\Delta E(\rho(\vec{r}))$~(see Eqn.\ref{eq:FermiShift}), and $\gamma$ is given by our 450\kHz~Fourier limited bandwidth.\\  

\begin{figure}[t!]
	\includegraphics[width = 1\columnwidth]{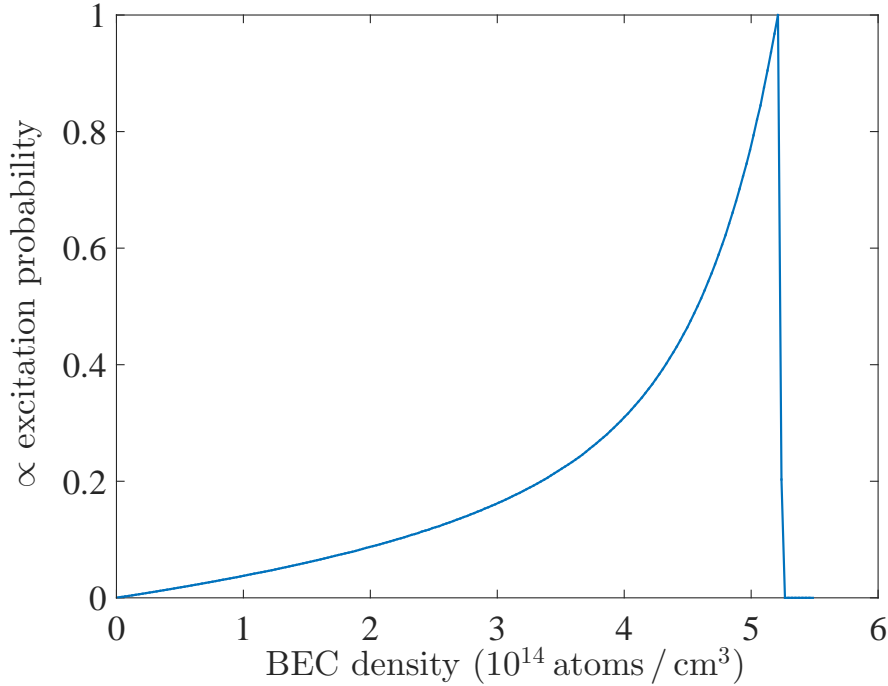}
	\caption{(color online) Excitation probability of a Rydberg atom in the BEC, according to equation~\ref{eq:scattering_rate}, considering a focused laser beam of 2.1\um~waist and an average peak density of \density{5.2}{14} realized over the first 50 excitation pulses. The excitation probability has its maximum at the peak density of the BEC because of the focused laser beam. 
}
\label{fig:effectivedensity}
\end{figure}

In Figure~\ref{fig:effectivedensity} the effective density spatial profile in a BEC is shown.  It is apparent from Figure~\ref{fig:effectivedensity} that due to the focused beam set-up in our experiment, we have a higher probability of exciting the highest density regions than if we illuminated the whole cloud.  This set-up therefore leads to Rydberg spectra in a quantum gas, which emphasize the signal in high densities.   

\subsection{Comparison of measurements}\label{Comparison}

\begin{figure}[t!]
	\includegraphics[width = 1\columnwidth]{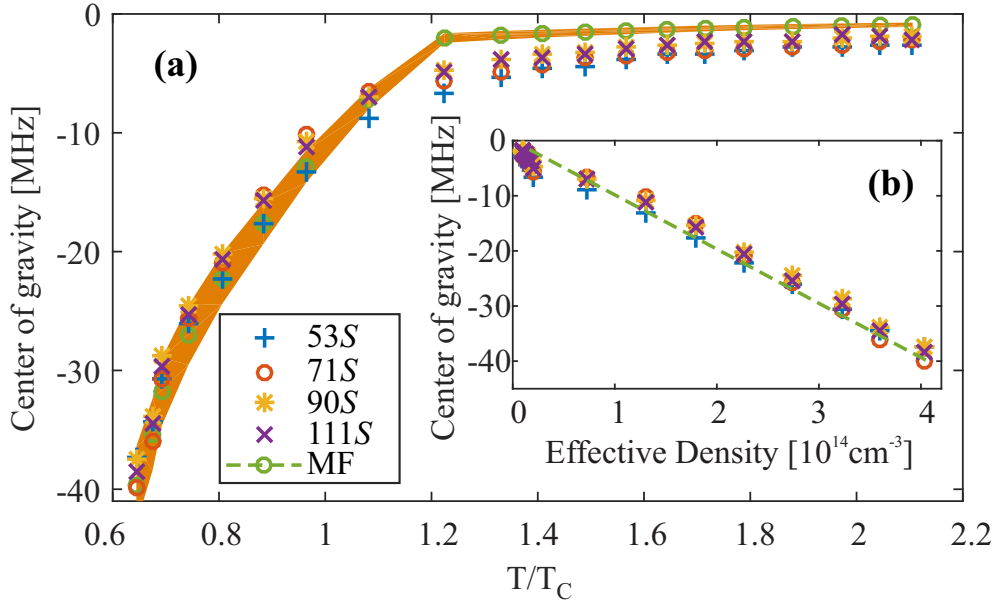}
	\caption{(color online) (a) The center of gravity for all measured states as a function of the temperature divided by the critical temperature (\textit{T}/\textit{T}$_{\rm{c}}$). The dashed line connects the predicted center of gravity shift using Fermi's mean-field (MF) model for the measured effective density described in section~(\ref{measuredDensity}).  The orange shaded region represents an error in the predicted Cog due to an estimated error in measured atom number of $\pm20\%$. (b) In the inset the Cog is plotted versus the effective density. The dashed line indicates again the center of gravity shift using Fermi's mean-field (MF) model for the measured effective density described in section(\ref{measuredDensity})}
	\label{fig:COG_Melting}
\end{figure}

From the spectra shown in Figure~\ref{fig:53SSpectra} we can extract the center of gravity (Cog) as a function of~\textit{T}/\textit{T}$_{\rm{c}}$ and effective density, displayed in Figures~\ref{fig:COG_Melting}(a) and~\ref{fig:COG_Melting}(b), respectively.  The average \textit{T}$_{\rm{c}}$ for the first block of 50 experimental pulses is 430\nK.  The Cog as a function of~\textit{T}/\textit{T}$_{\text{c}}$ shows a pronounced change in slope at \textit{T}\,=\,\textit{T}$_{\rm{c}}$, as expected by the phase transition from BEC to thermal cloud and the concurrent change in density with temperature. In Figure~\ref{fig:COG_Melting}(b) it can be seen that the Cog of the Rydberg spectra for all \textit{n} are linear with respect to the effective density within the excitation volume.  The linear slope fits well to the \textit{s}-wave triplet scattering length \textit{a}$_{\textit{T}\rm{,0}}$\,=\,-15.7\rbohr~\cite{Boettcher2015}.

The dashed green lines plotted in Figures~\ref{fig:COG_Melting}(a,b), connect calculated points by applying equation~(\ref{eq:FermiShift}) to the calculated effective density.  The orange shaded region in Figure~\ref{fig:COG_Melting}(b) represents an error bar for the green mean-field data points by using an estimated error in the detected atom number of $\pm20$\%.  An error in the atom number also causes an error in the x-axis of $\pm0.07$.  The Cogs for the spectra of~\textit{T}$>$\textit{T}$_{\rm{c}}$ are two times larger than expected from the measured effective density, and we have verified that potential detector saturation at larger signal levels in the thermal cloud is not the cause for this discrepancy.  We speculate that the disagreement is caused by non-equilibrium distributions or because the non-local wavefunction of the alkali atom perturber is not considered in Fermi's density shift model nor the point-particle perturber model. 

Using the point-particle perturber model as presented in section~(\ref{densityShift}), a density shift is calculated over the range of densities relevant to the bimodal density distribution of the atomic cloud used in the experiment.  This gives an effective density shift of~$\approx\,12$MHz/(\density{1}{14}) over the range of~\textit{n} investigated.  Even within the error of the effective density, given by an uncertainty of the atom number of $\pm20$\%, this simulated density shift does not fit well with the data.  Some of this discrepancy could be due to the limited scan range of the experimental spectra.  A limited scan range would, however, affect the Cog measurement for the 53\textit{S} spectra at the highest densities most significantly, and as expected from the simulation, the 53\textit{S} spectra do not have Cogs that deviate from the Cogs measured for \textit{n}\,$=$\,71, 90, and 111 over the range of densities examined in the experiment.  This discrepancy between the simulated line shift and the measured line shift is certainly of interest, since the point-particle perturber model predicts the line broadening so well.  In future experiments the validity of the point-particle perturber model in predicting the line shift could be tested by taking Rydberg spectra where deviations of the mean-field density shift are expected, e.g.~at lower~\textit{n}.             

\section{Localization of the Rydberg Excitation in the BEC}\label{Localization}

Evident from the point-particle perturber model introduced in section~\ref{densityShift}, a specific detuning of the excitation lasers selects particular configurations of background gas atoms with varying probabilities, as shown in Figure~\ref{fig:Configurations_vsn}.  It is worth investigating whether the Rydberg atom can reliably be excited with a given configuration or density of background gas atoms for an excitation laser with a particular bandwidth.  Reliably exciting the Rydberg atom with a specific background gas configuration also leads to a confined spatial placement of the Rydberg atom within the BEC due to the sharp density gradients of the Thomas Fermi profile.  

\subsection{Experimental Validity of Localization}\label{experimentLocalization}

To demonstrate the validity of localizing the Rydberg excitation for high \textit{n}, we perform the following experiment: We set a constant detuning of the infrared laser and we move the position of the BEC through the focused laser beam in the x-axis in steps of 900\nm, by changing the position of the magnetic trap along this axis.

\begin{figure}[t!]
	\includegraphics[width = 1\columnwidth]{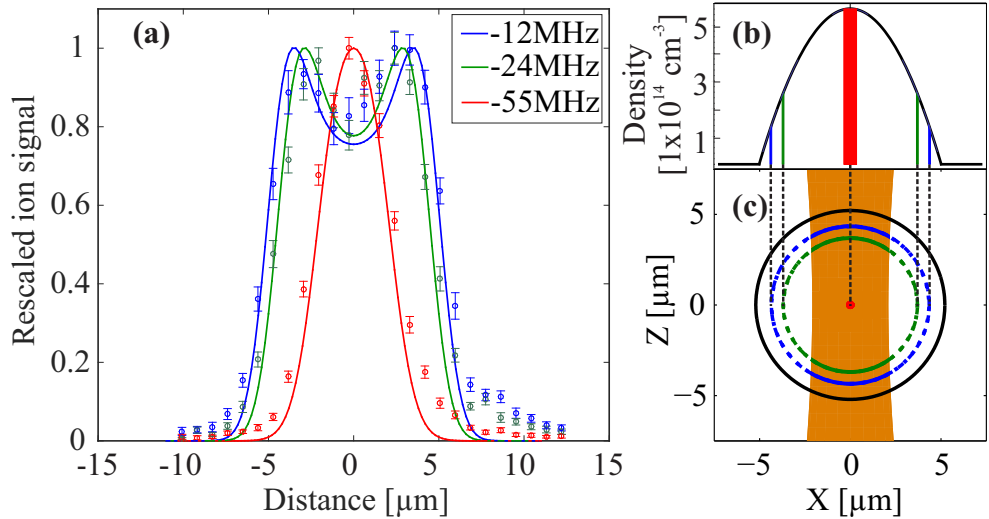}
	\caption{(color online)(a) The data (points with error bars) show the mean ion signal for the first 50 excitation pulses versus the position of the BEC with respect to the focused infrared laser beam, along the x-axis.  The lines represent the expected signal profile from simulations described in the text. (b) The schematic of the Thomas-Fermi profile with the lines indicating the probed density shells for the respective detunings. (c) Depiction of a cut through the BEC in the xz-plane. The solid part of the circle indicates the overlap between the Gaussian beam and the probed area of the density shell.}
	\label{fig:offset_scan}
\end{figure}

Figure~\ref{fig:offset_scan}(a) shows the results of these positional scans for detunings of -12, -24, and -55\MHz~for the 90\textit{S} state.  The signal width of these scans decreases with increased negative detuning as the region of the BEC, which is resonant with the excitation beams, reduces to the center of the BEC, as shown in~\ref{fig:offset_scan}(b).  A prominent dip is visible in the -12 and -24\MHz~scans when the excitation beams are not resonant with the central (highest-density) portion of the cloud; at these detunings we probe density shells shifted to the edges of the BEC spheroid.  The dip is only about 20\% of the total signal size as we integrate over the probed BEC distribution along the z-axis (see Figure~\ref{fig:offset_scan}(c)). \\

For the simulated spatial distributions shown in Figure~\ref{fig:offset_scan}(a) as solid lines, we use the basis of the spectra simulation described in section~\ref{densityShift} whereby the bimodal density profile, the assumed Gaussian 3D infrared beam shape and the spectral bandwidth of the excitation laser are known parameters in the model. The slight deviation of the data points from the simulations, for the highest detuning and at the region of more than 7\um~away from the BEC center, could be explained by our beam shape, which is not perfectly Gaussian. 

The agreement between data and the simulations indicates that for \textit{n}\,$=$\,90 we can restrict the Rydberg excitation, in the inhomogeneous density of the BEC, to shells with thicknesses given by the excitation bandwidth and density gradient, by controlling the detuning of the excitation lasers. For 2\us~excitation pulses, as used here, this corresponds to a spatial density shell thickness of about 50\nm~and 700\nm~with \textit{n}$_{\rm{low}}$\,=\,\density{1.5}{13} and \textit{n}$_{\rm{high}}$\,=\,\density{5.2}{14} at the edges and the peak of the Thomas-Fermi profile, respectively.  Furthermore, the data plotted in Figure~\ref{fig:offset_scan}(a) demonstrate that the density shell excited with -12\MHz~detuning, can be distinguished from the density shell excited with $-24$\MHz~detuning, although the density shells are only separated by $\approx$~600\nm.  For \textit{n}$\,=\,90S$, the size of the Rydberg electron wavefunction is $\approx$~1.6\um~in diameter, demonstrating that the center of gravity of the Rydberg wavefunction can be localized to length scales smaller than its size in a steep density gradient, if the spectral bandwidth is chosen accordingly.  If the second excitation laser is also focused, this technique could be used to characterize 3D profiles of an atomic cloud.

\section{Conclusion}\label{Conclusion}
In this review we discussed the original derivation of Fermi's density shift and subsequent extensions of the theory to higher-order scattering terms in order to more precisely describe the electron - alkali atom perturber scattering potentials.  We discussed the point-particle perturber model as a method to simulate the line shift and line broadening measured in high density Rydberg spectroscopy.  The model probes the two-body potential energy curves for particular background perturber configurations, rather than considering an average electron-perturber potential energy dominated by \textit{s}-wave scattering as in Fermi's density shift model. 
 
We studied the density-induced line shift in a \rubidium~BEC for single Rydberg excitations for a range of densities and temperatures.  We have confirmed experimentally a linear behavior of the density shift for the range of Rydberg states from 53\textit{S} to 111\textit{S}, in agreement with Fermi's mean-field density shift (see equation~(\ref{eq:FermiShift})) and the results of the point-particle perturber model for the density shift as shown in Figure~\ref{fig:DensityShift}.  This spectroscopy technique could be used to observe repeatable dynamics of the \insitu~density in a quantum gas on a fast time scale.  We have shown that Rydberg atoms can be localized within a density shell within the overlap of the excitation lasers for high principal quantum numbers.  In future work this technique could be employed for imaging the influence of the Rydberg electron on the density distribution of background perturber atoms~\cite{Karpiuk2015} and it was subsequently made use of for measuring collisional processes at various densities within the BEC~\cite{Schlagmueller2016PRX}. 

The point-particle perturber model performs well in simulating the line broadening of Rydberg spectra in a BEC~\cite{Schlagmueller2015}, and the localization of the Rydberg excitation shown in section~(\ref{Localization}), but less good agreement is found in simulating the magnitude of the density shift of high density Rydberg spectra.  It was found that the magnitude of the density shift for spectra taken in the BEC agreed well with the magnitude expected from Fermi's mean-field density shift, whereby only \textit{s}-wave scattering is considered.  The disagreement of the simulation in predicting the line shift of the experimental data beyond Fermi's mean-field density shift is intriguing.  It is unclear whether this disagreement is dominated by the inability of this semiclassical simulation to calculate the bound states of the PECs as discussed in section~\ref{densityShift}, or whether more exact potential energy curves are required as discussed in section~\ref{Resonance}, or whether there was an experimental error because of the limited scan range as discussed in section~\ref{Measurement}.      
Although the point-particle model does not demonstrate perfect agreement of the magnitude of the line shift, it could be used to predict anomalies of the mean-field density shift by considering the relevant potential energy curves of the system.  Modern cold atom experiments have increasing control over the species, spin, and wave function of the Rydberg and perturber atoms, which allows for investigation of the density shift for a large collection of potential energy curves.  

\indent We thank Stephan Jennewein, Christoph Tresp, and Udo Hermann for their contributions to the experiment.  We thank Georg Raithel for fruitful discussions.  The work is funded by the Deutsche Forschungsgemeinschaft (DFG) within the SFB/TRR21 and the project PF~381/13-1. This work is supported in part by the National Science foundation under Grant PHY-1306905.  S.H. is supported by the DFG through Project No. HO~4787/1-1. We also acknowledge support by the ERC under contract number 267100. H.N. acknowledges support from the Studienstiftung des deutschen Volkes. M.S. acknowledges support from the Carl Zeiss Foundation.
\indent
\bibliography{literaturenew_arXiv}
\end{document}